\documentclass{jaa}
\usepackage{natbib}

\usepackage{graphicx}

\def\msun{\hbox{M$_\odot$}}

\def\mnras{\hbox{MNRAS}}
\def\apj{\hbox{ApJ}}
\def\aj{\hbox{AJ}}
\def\aap{\hbox{A$\&$A}}

\begin{document}\sloppy

\title{On the dark matter formation scenario of NGC~4147}


\author{Andr\'es E. Piatti\textsuperscript{1,2,*}}
\affilOne{\textsuperscript{1}Instituto Interdisciplinario de Ciencias B\'asicas (ICB), 
CONICET-UNCUYO, Padre J. Contreras 1300, M5502JMA, Mendoza, Argentina; \\
\textsuperscript{2}Consejo Nacional de Investigaciones Cient\'{\i}ficas y T\'ecnicas 
(CONICET), Godoy Cruz 2290, C1425FQB,  Buenos Aires, Argentina}


\twocolumn[{

\maketitle

\corres{andres.piatti@fcen.uncu.edu.ar}


\begin{abstract}
We report results on the radial velocity dispersion profile built out to the outskirts of
NGC~4147, a Milky Way globular cluster with detected strong tidal tails.
The cluster was chosen to probe, from an observational point of view, recent simulations 
that suggest that rising velocity dispersion profiles at large distance from the clusters'
centers would be seen in globular clusters without tidal tails. From GEMINI@GMOS spectra,
centered in the infrared CaII triplet region, of selected stars located along the onset
of NGC~4147's tidal tails, we measured their radial velocities and overall metallicities.
The derived metallicities were used to ultimately assessing on the highly-ranked cluster candidates
of 9 stars, located between $\sim$ 7 and 33 pc from the cluster's center, suitable for
testing the aforementioned simulations. We complemented the present radial velocities with others
available in the literature for cluster's members, and built a cluster velocity dispersion
profile which suggests a mostly flat or slightly rising profile at large distances from the 
cluster's center. This outcome confirms that kinematically hot outermost cluster's stars are 
seen in NGC~4147, which disproves the recent model predictions. Nevertheless, the mean velocity 
dispersion of the outermost cluster's stars agrees with NGC~4147 being formed in a 
10$^8$-10$^9$$\msun$ dwarf galaxy with a cored dark matter profile that later was accreted on to the Milky Way.
\end{abstract}


}]



\section{Introduction}

Globular clusters are the oldest stellar systems formed when the Milky Way was born,
hence the importance of studies their formation scenario. Recent models by \citet{cg2021}
propose that globular clusters without tidal tails (relative long stellar streams 
that have escaped the cluster) formed within dark matter halos, which effectively 
cancel the tidal effects with the Milky Way which would have formed such tails
\citep{starkmanetal2020,boldrinietal2020,bv2021,wanetal2021}.
These models can be tested by simply checking for large velocity dispersion at large radial 
distance from the cluster center and comparing the overall velocity dispersion curve 
with various models with and without dark matter \citep{bonacaetal2019,cg2021}.
Up to date, none of the known Milky Way globular clusters have been targeted for building 
their velocity dispersion profiles at large distance from their centers
(r $>$ 3-6 times the half-mass radius), so that 
the above model remains untested. 

As a starting point, one could select globular clusters
from the stringent compilation of Milky Way globular clusters 
with robust detection of tidal tails \citep{zhangetal2022}. Their catalog includes 46 
globular clusters classified as follows: 27 with tidal tails, 4 with extended envelopes, 
and 15 without observed extended features. NGC~4147 is a globular cluster that shows one of 
the strongest tidal tails \citep{jg2010}, so that it is a suitable defiant candidate to test the predicted 
formation scenario  of globular clusters within dark matter halos, from the construction 
of its velocity dispersion profile at large distances and the comparison with formation 
models with and without dark matter. NGC~4147 has been associated to the Helmi streams group
\citep{massarietal2019,malhanetal2022b,callinghametal2022} and to the Gaia-Sausage-Enceladus galaxy \citep{zhangetal2024}.

We note  that the aforementioned formation models need 
to be tested from a statistical significant globular cluster sample with and without tidal 
tails. Nevertheless, we perform here a first exploration of the testing procedure of these 
formation models. To this respect, we prefer to target a globular cluster with strong tidal 
tails like NGC~4147 to rule out rising 
velocity dispersion profiles, rather than targeting globular clusters without tidal tails to 
confirm rising profiles, because the latter does not have stars in its outskirts to 
perform the experiment. As a guidance, \citet[][see Figure~12]{cg2021} show the most promising 
velocity dispersion profile of NGC~4147 obtained up-to-date. They overplotted  possible 
profiles for different formation models, namely:  a W=7 \citet{king62}'s model (red); a 
simulated globular cluster in the Milky Way dark matter background (green), and a cluster 
at its sub-halo center  (blue). As can be seen, without data beyond $\sim$ 12 pc, it is not 
possible to know the model that best resembles the velocity dispersion profile in the 
outskirts of NGC~4147.

Precisely, we embarked in an observational program with the aim of obtaining radial velocities
of NGC~4147's red giants located from nearly 7 pc up to 35 pc 
from the cluster's center, just covering the distance range  necessary to probe the aforementioned 
formation models. In Section~2, we present the acquired observational data, and in Section~3 
we derive different astrophysical properties of the target stars. Section~4 is devoted to the 
discussion on NGC~4147's radial velocity dispersion profile at large distance from its center. 
Finally, Section~5 summarizes the main conclusions of this work.

\begin{figure}
\includegraphics[width=\columnwidth]{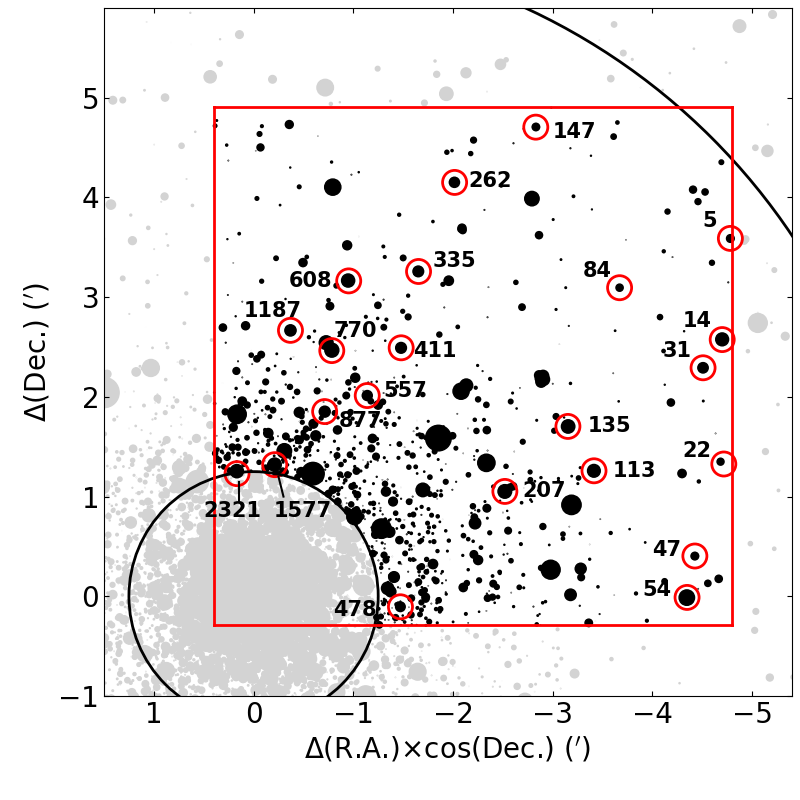}
\caption{GMOS FOV (red square) centered on the studied region. Gray filled circles are 
\citet{stetsonetal2019}'s $UBVRI$ photometric catalog of stars, while black filled circles 
highlight those for the region of interest; the size of the symbol is proportional to the
Johnson $I$ magnitude. Large red open circles are
drawn for stars with obtained spectra. Black circles of radius 7 and 35 pc, respectively,
are also drawn.}
\label{fig1}
\end{figure}

\begin{figure}
\includegraphics[width=\columnwidth]{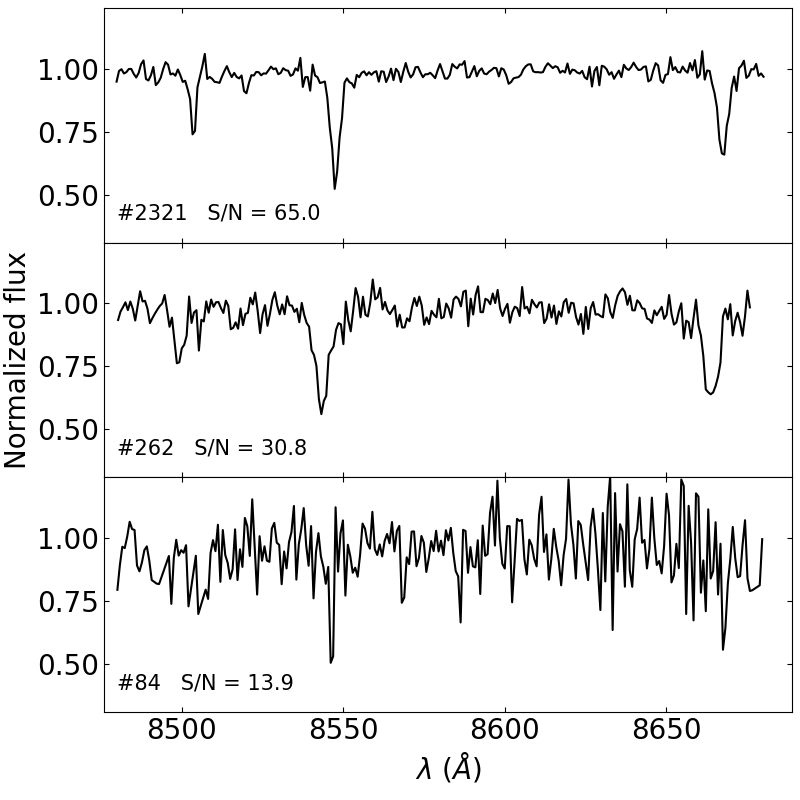}
\caption{Normalized spectra of some observed stars.}
\label{fig2}
\end{figure}

\section{Observational data}

With the aim of selecting the most promising stars for measuring the dispersion
velocity in the cluster's outskirts, we focused on the onset of the cluster's tidal tail
that points to the Northern-Northwestern direction from the NGC~4147's center 
\citep[see Figure~9 in][]{jg2010}. We considered the FOV of the GEMINI Multi-Object Spectrograph 
(GMOS, $\sim$ 5.5'$\times$ 5.5'), the awarded observing time (program GN-2024A-FT-212, 
PI: Piatti), the stellar distribution in the region according to our Johnson $I$ preimage, 
and the ability to add as many individual 
slits as possible in a single manufactured MOS mask. We did not consider proper motions as a criterion 
for selecting stars, since the target candidates could not share the mean cluster proper motions.
Instead, we assigned final cluster membership status from the derived Ca~II triplet metallicity
(see below). NGC~4147 is located near the North galactic pole ($\it l$= 252.84$^o$, b=+77.18$^o$),
so that Milky Way field star contamination is not a significant constraint. Figure~\ref{fig1}
illustrates the spatial distribution in the Johnson $I$ pass-band of the selected stars (marked 
with a large red open circle) 
across the onset of the Northern cluster's 'tidal tail, just covering the distance range 
necessary to probe the aforementioned formation models.

We used GMOS with the mid-resolution (R $\approx$ 2200) R831 grism centered on the Ca~II triplet
($\sim$ 8500$\AA$) with the OG515 blocking filter ($>$ 5200 $\AA$), which gives a dispersion of 0.76 
$\AA$/pixel for the selected binned pixel
configuration. We obtained a total exposure of 2700 sec for the observing science field 
(Figure~\ref{fig1}) using slit widths of 1.0" placed on the selected stars. The slit length allowed
individual background sky subtraction during the data processing. The observational material
was reduced with the dedicated IRAF@GEMINI.GMOS package following the instructions provided by
the GEMINI Observatory\footnote{https://nsf-noirlab.gitlab.io/csdc/usngo/gmos-cookbook/}.
These tasks include preparing the data for the reduction; processing and combining 
bias and flats: applying overscan, trimming, mosaicing, bias and flats; performing wavelength 
calibration with CuAr arc spectra taken before and after the science exposures; 
rectifying and wavelength calibrating the MOS exposures; and extracting the sky-subtracted individual 
star spectra. Table~\ref{tab1} lists the observed stars
alongside their equatorial coordinates, the \citet{stetsonetal2019}'s Johnson $V$ magnitude, 
and  average signal-to-noise ratio (S/N) along the whole wavelength range of each spectrum. In 
Figure~\ref{fig2} we show some of the spectra spanning most of the obtained S/N ratio range.

\section{Stellar parameter estimates}
\subsection{Radial velocity measurements}

We measured radial velocities (RVs) by cross-correlating the observed spectra 
and synthetic ones taken from the PHOENIX library\footnote[1]{http://phoenix.astro.physik.uni-goettingen.de/} 
\citep{husser2013}, following the procedure described in \citet{piattietal2018c}. In brief, 
we selected templates with $T_{\textrm{eff}}$ in the range 4500-5500 $^{\rm o}$K and log~$g$ 
between 2.5-3.5 dex, which correspond to giant stars with MK types $\sim$ G0-K4. We restricted  
the templates to those with the mean cluster metallicity 
\citep[{\rm [}Fe/H{\rm ]}=-1.81 dex,][]{carrettaetal2009,villanovaetal2016}.
All the spectra were continuum normalized before the cross-correlation procedure and 
the synthetic spectra resolution was degraded to match the resolution of our
science spectra. Each observed spectrum was cross-correlated against the whole selected 
synthetic template sample by making use of the IRAF@FXCOR task, which implements the algorithm 
described in \citet{tonry1979} for the construction of the cross-correlation function of each 
(object, synthetic) spectra pair. In addition to the RVs,  FXCOR returns the CCF normalized peak 
($h$), which is an indicator of the degree of similarity between the correlated spectra and the 
Tonry \& Davis ratio (TDR) defined as TDR$=h/(\sqrt{2}\sigma_{a})$, where $\sigma_{a}$ is root mean 
square of the CCF antisymmetric component. For each object spectrum we assigned the RV value 
resulting from the cross-correlation with the highest $h$ value, which was in all cases greater 
than 0.8. Finally, we carried out the respective heliocentric corrections by using the 
IRAF@RVCORRECT task. Table~\ref{tab1} lists the resulting heliocentric RVs with their respective 
uncertainties. 

\begin{table}
\caption{RVs of the studied star sample.}
\label{tab1}
\begin{small}
\begin{tabular}{lccccr}\hline\hline
ID     &  R.A. & Dec.         & $V$  & $<$S/N$>$ &  RV\hspace{0.7cm}    \\ 
       &     (hr)  & (dec)    & (mag)    &            &  (km/s)\hspace{0.5cm} \\\hline
    5  &  12.1628 & 18.602 &  20.20  & 20.9       & -69.30$\pm$2.05\\        
   14  &  12.1629 & 18.585 &  18.59  & 55.6       &  17.29$\pm$2.89\\  
   22  &  12.1629 & 18.565 &  21.03  & 10.2       & 168.95$\pm$2.71\\     
   31  &  12.1631 & 18.580 &  19.34  & 31.3       & -92.22$\pm$3.82\\        
   47  &  12.1632 & 18.549 &  20.11  & 18.7       & 183.55$\pm$1.33\\     
   54  &  12.1633 & 18.542 &  19.90  &114.5       &  -5.20$\pm$2.12\\      
   84  &  12.1641 & 18.594 &  20.25  & 13.9       & 183.05$\pm$1.67\\  
  113  &  12.1644 & 18.563 &  18.74  & 47.7       & 181.94$\pm$2.97 \\       
  135  &  12.1647 & 18.571 &  19.40  & 59.6       & -17.09$\pm$2.32\\       
  147  &  12.1651 & 18.621 &  21.83  & 19.3       &  15.92$\pm$1.82\\     
  207  &  12.1654 & 18.560 &  18.46  & 60.1       & 181.49$\pm$2.29\\   
  262  &  12.1660 & 18.612 &  20.21  & 30.8       &  11.06$\pm$3.14\\            
  335  &  12.1664 & 18.597 &  20.78  & 48.8       &  67.95$\pm$3.46\\         
  411  &  12.1666 & 18.584 &  19.33  & 38.3       & 182.38$\pm$1.69\\        
  478  &  12.1667 & 18.540 &  19.68  & 28.1       & 178.42$\pm$2.35 \\       
  557  &  12.1670 & 18.576 &  19.56  & 28.4       & 180.82$\pm$2.25\\          
  608  &  12.1673 & 18.595 &  18.89  & 61.0       & -82.31$\pm$2.71\\          
  770  &  12.1675 & 18.583 &  19.97  & 71.1       &  27.24$\pm$3.93\\         
  877  &  12.1675 & 18.573 &  19.28  & 37.3       & 172.90$\pm$5.25\\           
 1187  &  12.1679 & 18.587 &  19.20  & 39.8       & 177.73$\pm$3.01\\   
 1577  &  12.1681 & 18.564 &  18.58  & 29.3       & 144.63$\pm$3.81\\           
 2321  &  12.1686 & 18.563 &  20.79  & 65.0       & 174.94$\pm$1.83\\\hline
\end{tabular}
\noindent Note: NGC~4147's mean RV = 179.35$\pm$0.31 km/s \citep{bh2018}.
\end{small}
\end{table}

\subsection{Overall metallicity measurements}

\begin{figure}
\includegraphics[width=\columnwidth]{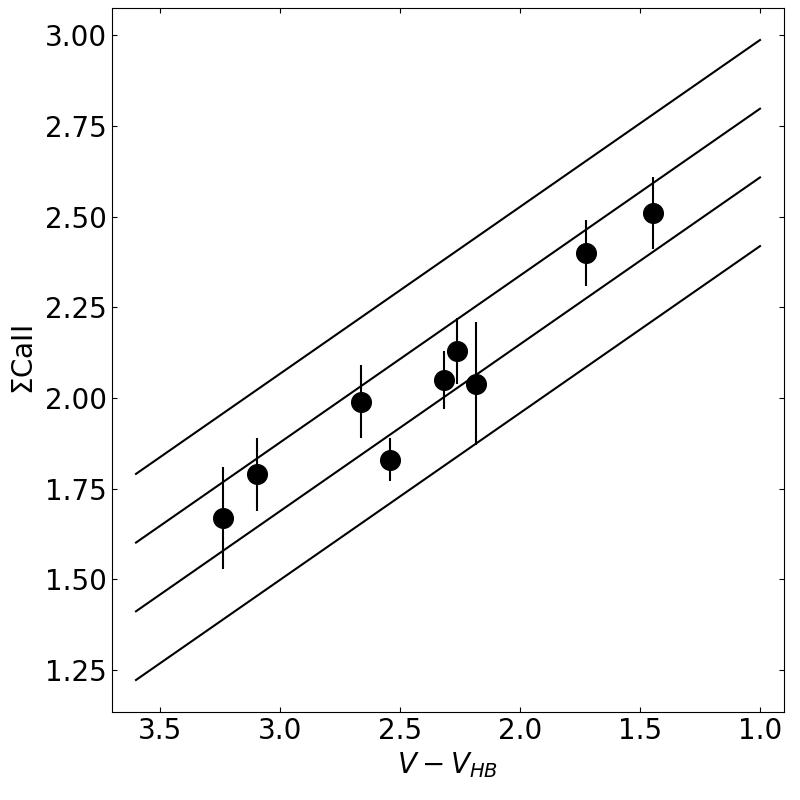}
\caption{Sum of the CaII triplet line equivalent widths, included
error bars,  as a function
of $V-V_{BH}$ for stars observed in the field of NGC~4147.
The isoabundance lines for [Fe/H] = -1.60 dex down to -1.90 dex
(top to bottom) in steps of $\Delta$[Fe/H] = 0.10 dex are also overplotted.}
\label{fig3}
\end{figure}

Equivalent widths ($W$) of the infrared CaII triplet lines were measured from the normalized
spectra using the SPLOT package within IRAF. Since \citet{cg2021} showed that the
velocity dispersion of NGC~4147's stars could rise up to $\sim$ 15 km/s in the observed 
distance range from the cluster's center, we constrained the CaII measurements to those stars
with RVs within $\pm$30 km/s from the mean cluster's RV \citep[179.35$\pm$0.31 km/s,][]{bh2018}.
We defined different local  continua over each spectral line, bearing in mind the presence 
of TiO bands and the spectra S/N ratio, in order to estimate the corresponding equivalent
width uncertainties. The resulting values are listed in Table~\ref{tab2}. We then overplotted 
the sum of the equivalent widths of the two stronger CaII lines ($\Sigma$$W$(CaII) = 
$W$(Ca8542$\AA$) + $W$(Ca8662$\AA$)) in the 
$\Sigma$W(CaII) versus $V-V_{\rm HB}$ plane, that has been calibrated in terms of 
metallicity \citep[see, ][]{dacosta2016}. In that diagram $V_{\rm HB}$ refers
to the mean magnitude of the globular cluster horizontal branch. We adopted a $V_{HB}$ values of 
17.02 mag, taken from the \citet[][2010 Edition]{harris1996}'s catalog of parameters for globular 
clusters in the Milky Way, while the individual $V$ magnitudes were taken from the $UBVRI$ 
photometric catalog of \citet{stetsonetal2019} (see Table~\ref{fig1}).  Figure~\ref{fig3}
depicts the resulting relationship. From $\Sigma$$W$(CaII), we computed the reduced equivalent 
width $W$',  which is defined as the value of $\Sigma$$W$(CaII) at $V-V_{HB}$ = 0.0 mag. $W$' 
was calibrated in terms of [Fe/H] by \citet{dacosta2016}, which is illustrated with isoabundance 
lines in Figure~\ref{fig3}. They correspond, from top to bottom, to [Fe/H] = -1.60 dex down to 
-1.90 dex, in step of $\Delta$[Fe/H] = 0.10 dex. Finally, we interpolated Figure~\ref{fig3} to 
obtain the individual overall metallicities and their respective propagated uncertainties 
(see Table~\ref{tab2}).

\begin{table}
\caption{CaII measures of some studied stars.}
\label{tab2}
\begin{small}
\begin{tabular}{lccc}\hline\hline
ID     &    W8542  & W8662 & [Fe/H]  \\ 
       &    (\AA) &  (\AA)  &  (dex) \\\hline
  22   &    1.40$\pm$0.04 &  1.00$\pm$0.06 & -1.18$\pm$0.05 \\
  47   &    0.90$\pm$0.05 &  0.89$\pm$0.05 & -1.72$\pm$0.05 \\
  84   &    0.92$\pm$0.05 &  0.75$\pm$0.09 & -1.75$\pm$0.07 \\
  113  &    1.30$\pm$0.05 &  1.10$\pm$0.04 & -1.73$\pm$0.05 \\
  207  &    1.28$\pm$0.07 &  1.13$\pm$0.03 & -1.74$\pm$0.05 \\ 
  411  &    1.08$\pm$0.13 &  0.97$\pm$0.05 & -1.77$\pm$0.04 \\
  478  &    1.23$\pm$0.06 &  0.76$\pm$0.04 & -1.72$\pm$0.05 \\
  577  &    1.16$\pm$0.04 &  0.67$\pm$0.02 & -1.84$\pm$0.03 \\
  877  &    1.31$\pm$0.02 &  0.82$\pm$0.17 & -1.75$\pm$0.05 \\ 
 1877  &    1.17$\pm$0.20 &  0.87$\pm$0.08 & -1.81$\pm$0.09 \\
 2321  &    1.25$\pm$0.09 &  0.91$\pm$0.11 & -1.36$\pm$0.06 \\\hline
\end{tabular}

\noindent Note: NGC~4147's mean [Fe/H] = -1.78$\pm$0.08 dex \citep{carrettaetal2009}.
\end{small}
\end{table}

\section{Analysis and discussion}

Nine of the eleven stars with CaII tiplet metallicities (see Table~\ref{tab2})
show very similar values, with an average of [Fe/H] = -1.76$\pm$0.04 dex. 
We considered them as highly-ranked cluster member candidates, as judged
by comparing their metallicities with the mean cluster's values 
derived from high-dispersion spectroscopy studies, e.g.: [Fe/H] = -1.78$\pm$0.08
dex \citet{carrettaetal2009} and -1.84$\pm$0.02 dex \citep{villanovaetal2016}.
These stars are located between $\sim$7 and 32 pc from the cluster's
center, so that they are valuable for building for the first time the velocity 
dispersion profile of NGC~4147 out to its outskirts.

\citet{cg2021} used {\it Gaia} DR3 proper motions \citep{gaiaetal2016,gaiaetal2022b} 
of selected stars filtered by their position in the cluster's color-magnitude diagram
to conclude that they did find no stars beyond 3 times the 3D half-mass radius, 
a distance from the cluster's center from which is possible to differentiate between 
velocity dispersion profiles of NGC~4147 following a \citet{king1966}'s model, a simulated 
cluster in the Milky Way dark matter background, and a cluster at its sub-halo
center. 

We here used the derived RVs of the nine stars mentioned above (see also 
Table~\ref{tab1}) and those previously obtained by \citet{bh2018} for 42 stars with
membership probabilities higher than 90$\%$ to construct the cluster velocity dispersion
profiles along nearly 77 pc from the cluster's center We doubled the number of stars
located between $\sim$ 7 and 32 pc, and used 51 stars in total, four of them taken from
\citet{bh2018} located between $\sim$ 32 and 77 pc from the cluster's center.  

\begin{figure}
\includegraphics[width=\columnwidth]{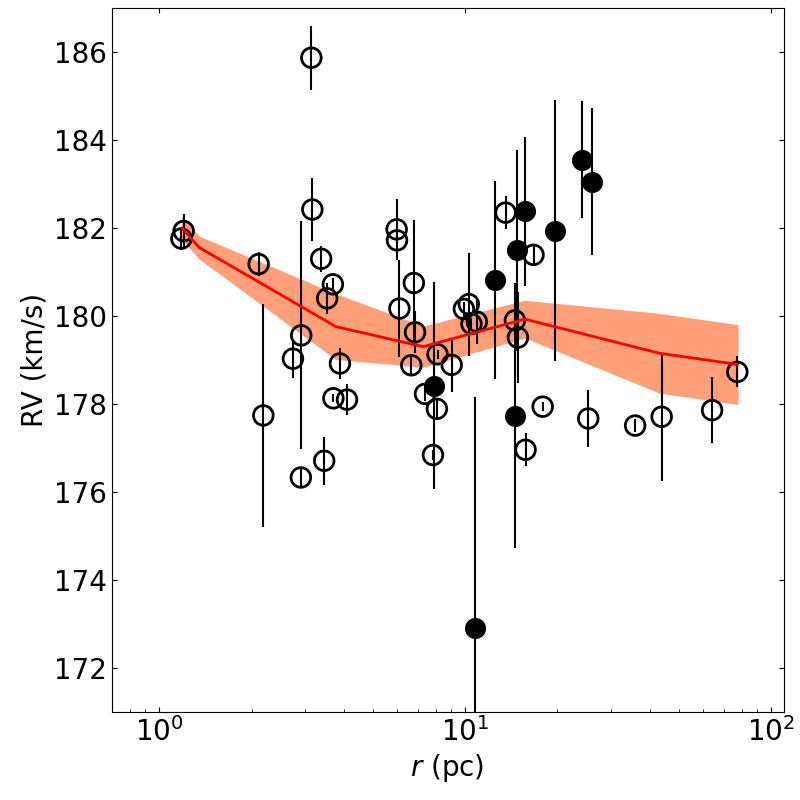}
\caption{RVs and their respective error bars of uncertainty as a function 
of the distance from the cluster's center for stars measured by \citet{bh2018} and
in this work, represented by open and filled circles, respectively. The solid
line and its shaded region represent the mean value and its standard deviation,
respectively.}
\label{fig4}
\end{figure}

\begin{figure}
\includegraphics[width=\columnwidth]{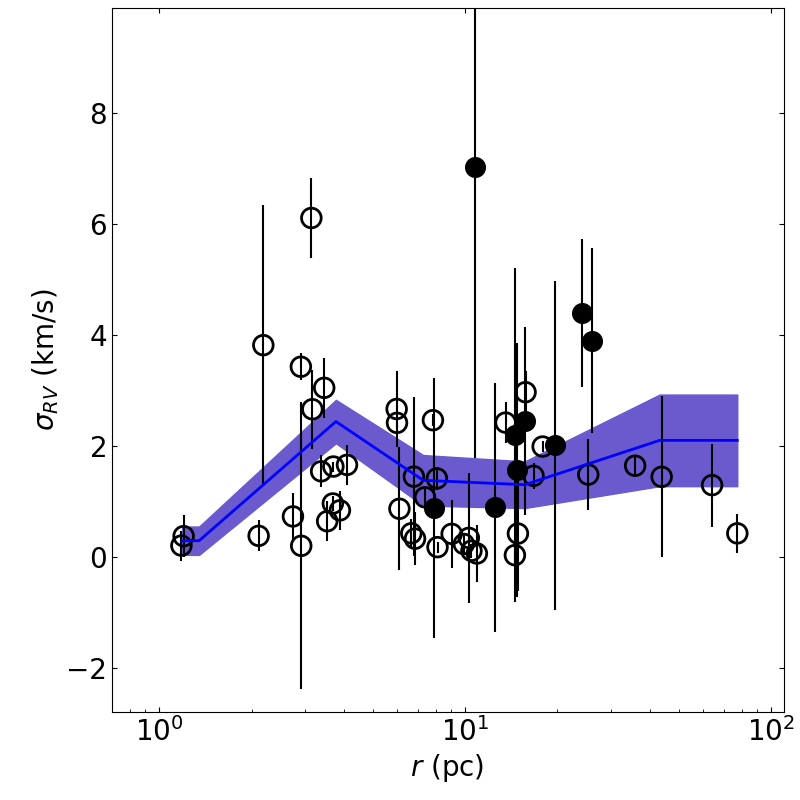}
\caption{Same as Figure~\ref{fig4} for the velocity dispersion (see text for
detals).}
\label{fig5}
\end{figure}

We started by determining the mean RV and dispersion for different distance bins (see 
Table~\ref{tab3}) by
employing a maximum likelihood approach, which accounts for each individual star's measurement 
errors that could artificially inflate the dispersion if ignored. 
We optimized the probability $\mathcal{L}$ that a given ensemble of stars  (N) with 
radial velocity RV$_i$ and radial velocty errors $\sigma_i$ are drawn from a population with mean
radial velocity $<$RV$>$ and intrinsic dispersion $W$  \citep[e.g.,][]{walkeretal2006,Koch2018Gaia}, 
as follows:

\begin{eqnarray*}
\mathcal{L}\,=\,&\prod_{i=1}^N&\, \left(\,2\pi\left[\sigma_i^2 + W^2 \, \right]\,\right)^{-\frac{1}{2}} \\
&\times&\,\exp \left(-\frac{\left({\rm RV}_i \,- <{\rm RV}]>\right)^2}{\sigma_i^2 + W^2} \right), 
\end{eqnarray*}

\noindent where the errors on the mean and dispersion were computed from the respective covariance 
matrices.  Table~\ref{tab3} lists the resulting mean RVs and RV dispersion for different ranges of
distances from the cluster's center. From these results, we built Figure~\ref{fig4}, which illustrates
the spatial distribution of the considered RVs and the corresponding mean values with their
standard deviations. Finally, we drew in Figure~\ref{fig5} the RV residual (individual value - mean 
value for the respective distance from the cluster's center) and superimposed the resulting $W$ values.
As can be seen, NGC~4147 would seem to have a flat or slightly rising velocity dispersion profile
toward the cluster's outskirts, with an average value of $\sigma$$_{RV}$($r$$>$20 pc) = 2.1$\pm$0.8 km/s.

\begin{table}
\caption{RV dispersion profile of NGC~4147.}
\label{tab3}
\begin{tabular}{cccc}\hline\hline
r        &  $<$RV$>$        & $W$    & $N^*$  \\ 
(pc)     &  (km/s)          & (km/s) &  \\\hline
$<$2.7   &  181.56$\pm$0.24 & 0.29$\pm$0.25 & 4 \\
2.7-4.8  &  179.76$\pm$0.73 & 2.44$\pm$0.39 & 12\\
4.8-9.7  &  179.30$\pm$0.45 & 1.38$\pm$0.45 & 12 \\
9.7-21.5 &  179.92$\pm$0.41 & 1.30$\pm$0.42 & 16 \\
$>$21.5  &  179.15$\pm$0.89 & 2.10$\pm$0.82 & 7 \\\hline
\end{tabular}

\noindent Note: $`*$number of stars used per bin.
\end{table}

The present outcome would not seem to favor a dark matter formation scenario 
for NGC~4147 as modeled by \citet{cg2021}. However, the resulting $\sigma$$_{RV}$
is in very good agreement with the cluster's speculated origin associated either
to the Helmi streams group \citep{callinghametal2022} or to the Gaia-Sausage-Enceladus 
galaxy \citep{zhangetal2024}. \citet{malhanetal2021} proposed that the dispersion in
line-of-sight velocity of globular cluster tidal tails tell us about the origin of Milky 
Way globular clusters, namely, whether they were accreted or formed in-situ. 
Globular clusters formed in a low-mass galaxy halo with cored or cuspy central density profiles 
of dark matter that later merged with the Milky Way develop tidal tails with different mean 
values of the velocity dispersion. On average, globular clusters from cuspy profiles have 
tidal tails with line-of-sight velocity dispersion being three times larger than those of clusters 
formed in cored dark matter profiles. Globular clusters formed in-situ have mean values of 
line-of-sight velocity dispersion nearly ten times smaller than those for globular clusters 
accreted inside  cored subhalos. Particularly, for streams formed in-situ, 
\citet{malhanetal2021} found line-of-sight velocity dispersion  $<$ 1 km s$^{\rm -1}$; 
for 10$^8$ and 10$^9$$\msun$ 
cored dark matter profiles ($\pm$3$\sigma$) $\sim$1-2 km s$^{\rm -1}$ and $<$5 km s$^{\rm -1}$, 
and for  10$^8$ and 10$^9$$\msun$ cuspy dark matter profiles  ($\pm$3$\sigma$) $\sim$7-12 km 
s$^{\rm -1}$. As can be seen, our mean $\sigma$$_{RV}$ value for the outskirts of NGC~4147 is 
consistent with a 10$^8$-10$^9$$\msun$ cored dark matter profile, which agrees with the expected stellar mass 
of Gaia-Sausage-Enceladus galaxy and Helmi streams \citep{callinghametal2022}, the two
associated origins claimed in the literature.

\section{Conclusions}

Recently, \citet{cg2021} proposed that the region between $\sim$ 3 and 6 times the
3D half-mass radius of the velocity dispersion profile of a Milky Way globular cluster 
could help to disentangling between different globular cluster formation scenarios, such
as a formation without dark matter, a formation in the Milky Way dark matter background,
or a formation in a dark matter sub-halo. These formation models would imprint a 
falling, a flat, or a rising velocity dispersion profile, respectively.

We here probed these models by building for the first time the radial velocity dispersion
profile out to the outskirts of NGC~4147, a Milky Way halo globular cluster with strong tidal 
tails \citep{jg2010}. 
We obtained intermediate-dispersion spectra in the region of the infrared CaII triplet of a selected 
sample of stars distributed in the cluster's outskirts, along the onset of its tidal tails, in order 
to have a good statistics. From the spectra, we derived individual stellar radial velocities
and overall metallicities; the latter being useful to ultimately assessing on the highly-ranked 
candidates to have been formed in NGC~4147. Finally, we confirmed 9 stars located between
$\sim$ 7 and 32 pc from the cluster's center with [Fe/H] values
in excellent agreement with the mean high-dispersion metallicity of NGC~4147. 

The resulting radial velocity dispersion profile, which included published RVs for other cluster's
members suggests a mostly flat or slightly rising profile at large distances from the cluster's center,
with a mean value of 2.1 km/s. The proposed models by \citet{cg2021} would not seem to match the
present findings, in the sense that rising velocity dispersion profiles would be linked to globular
clusters without tidal tails, which is not the case of NGC~4147. However, according to the models by
\citet{malhanetal2021}, NGC~4147 could have been formed in a 10$^8$-10$^9$$\msun$ dwarf galaxy with a cored
dark matter profile that later was accreted on to the Milky Way. This origin is in very
good agreement with two different possible observational-based associated progenitors.

\section*{Acknowledgements}
We thank the referee for the thorough reading of the manuscript and
timely suggestions to improve it.

Data for reproducing the figures and analysis in this work will be available upon request
to the author.


\end{document}